\documentclass[10pt]{iopart}
\usepackage{iopams}

\usepackage{graphicx}
\usepackage{color}
\usepackage[english]{babel}
\usepackage[pdfborder={0 0 0}, colorlinks=true, linkcolor=blue, citecolor=blue, urlcolor=cyan]{hyperref}

\def\beq{\begin{equation}}
\def\eeq{\end{equation}}
\def\bea{\begin{eqnarray}}
\def\eea{\end{eqnarray}}

\def\ra{\rangle}
\def\la{\langle}

\def\q{\mathbf{q}}

\def\S{\mathbf{S}}

\newcommand{\ii}{{\mathrm{i}}}

\newcommand{\pp}{{++}}

\newcommand{\Tmat}{\mathcal{T}}

\newcommand{\kv}{{\mathbf{k}}}

\begin{document}

\title[Decisive proofs of the $s_\pm \to s_{++}$ transition in $\lambda_L(T)$]{Decisive proofs of the $s_\pm \to s_{++}$ transition in the temperature dependence of the magnetic penetration depth}

\author{V.A. Shestakov$^1$, M.M. Korshunov$^{1,2}$, Yu.N. Togushova$^2$, and O.V. Dolgov$^{3,4}$}
\address{$^1$Kirensky Institute of Physics, Federal Research Center KSC SB RAS, 660036, Krasnoyarsk, Russia}
\address{$^2$Siberian Federal University, 660041, Krasnoyarsk, Russia}
\address{$^3$P.N. Lebedev Physical Institute RAS, 119991, Moscow, Russia}
\address{$^4$Donostia International Physics Center, 20018, San Sebastian, Spain}
\ead{mkor@iph.krasn.ru}

\vspace{10pt}
\begin{indented}
\item[]\date{\today}
\end{indented}

\begin{abstract}
One of the features of the unconventional $s_\pm$ state in iron-based superconductors is possibility to transform to the $s_{++}$ state with the increase of the nonmagnetic disorder. Detection of such a transition would prove the existence of the $s_\pm$ state. Here we study the temperature dependence of the London magnetic penetration depth within the two-band model for the $s_\pm$ and $s_{++}$ superconductors. By solving Eliashberg equations accounting for the spin-fluctuation mediated pairing and nonmagnetic impurities in the $T$-matrix approximation, we have derived a set of specific signatures of the $s_\pm \to s_{++}$ transition: (1) sharp change in the behavior of the penetration depth $\lambda_{L}$ as a function of the impurity scattering rate at low temperatures; (2) before the transition, the slope of
$\Delta \lambda_{L}(T) = \lambda_{L}(T)-\lambda_{L}(0)$ increases as a function of temperature, and after the transition this value decreases; (3) the sharp jump in the inverse square of the penetration depth as a function of the impurity scattering rate, $\lambda_{L}^{-2}(\Gamma_a)$, at the transition; (4) change from the single-gap behavior in the vicinity of the transition to the two-gap behavior upon increase of the impurity scattering rate in the superfluid density $\rho_{s}(T)$.
\end{abstract}

\pacs{74.20.Rp,74.25.-q,74.62.Dh}

\vspace{2pc}
\noindent{\it Keywords}: unconventional superconductors, iron pnictides, iron chalcogenides, impurity scattering, penetration depth

\submitto{\SUST}

%
\ioptwocol

\section{Introduction}

Unconventional superconductivity is full of surprises even in the thought to be simple cases. For example, $T_c$ in conventional superconductors with the $s$-wave gap is insensitive to the nonmagnetic disorder and decreases rapidly with the increasing number of magnetic impurities~\cite{Anderson1959,AGeng}. On the contrary, the unconventional sign-changing $d$-wave gap in the high-$T_c$ cuprates and $s_\pm$ gap in the iron-based superconductors leads to the suppression or in some cases to the saturation for the $s_\pm$ gap) of the critical temperature by scattering on the nonmagnetic impurities~\cite{Morozov1979,Preosti1994,Golubov1995,Kulic1999,Arseev2002,Balatsky2006,Scheurer2015}. Even more fascinating is the possibility to change the gap structure of multiband superconductors by the disorder. That is, the nodal angular-dependent extended $s$-wave gap may become nodeless due to the impurity averaging~\cite{v_mishra_09}. And the $s_\pm$ gap may transform to the sign-preserving $s_{++}$ gap with the increase of disorder~\cite{EfremovKorshunov2011,KorshunovUFN2016,ShestakovKorshunovSUST2018}. Even the reverse transition, from the $s_{++}$ to the $s_\pm$ state, is possible when the effect of temperature is considered~\cite{ShestakovKorshunovSymmetry2018}. Existence of such an interesting $s_\pm \leftrightarrow s_{++}$ transition, however, have to be proved. Observation of it may be tricky because the both states are fully gapped and both order parameters belong to the same symmetry class. Thus the specific heat and thermal conductivity measurements as well as angle-resolved photoemission spectroscopy (ARPES) are not the decisive tools of the first choice because they do not provide a direct probe of the changes happening at the transition. Intriguing possibility comes from the temperature dependence of the optical response and the London penetration depth $\lambda_L(T)$. In particular, the $s_\pm \leftrightarrow s_{++}$ transition goes through the gapless state that should reveal itself as the qualitative change of the form of $\lambda_L(T)$~\cite{EfremovKorshunov2011,KorshunovUFN2016}.

There's been at least two reports claiming the observation of the discussed transition. Namely, in Ba(Fe$_{0.9}$Co$_{0.1}$)$_2$As$_2$~\cite{Schilling2016} and in Ba(Fe$_{1-x}$Rh$_x$)$_2$As$_2$~\cite{Ghigo2018} with the nonmagnetic disorder in both systems introduced via a proton irradiation. Latter study is supplied with the theoretical calculations, which are questionable in some points. Firstly, plasma frequency $\omega_p$ changes with disorder by an order of magnitude, from 1.21~eV to 0.0821~eV and then grows back to 0.139~eV\footnote{Refs.~\cite{Torsello2019,Torsello2019CaKFeAs} and supplementary materials for Ref.~\cite{Ghigo2018} contains claim that $\omega_p$ is of the order of a few meV's, which is a typo since the typical value for metals should be around few eV's.}. Since $\omega_p$ is the property of the conduction electrons and there is no evidences that the proton irradiation severely affects the Fermi surface, there is no grounds to expect such a huge change in $\omega_p$.
The same is true for the analysis in the subsequent studies by the same group~\cite{Torsello2019,Torsello2019CaKFeAs}. Secondly, generalized cross-section parameter $\sigma$ also changes very much with the irradiation, from 0 to 0.278. The parameter itself controls the considered scattering limit, i.e. it equals zero in the weak (Born) limit and becomes unity in the strong (unitary) limit. Again, there is no solid grounds to expect that $\sigma$ would change with the addition of disorder created by the protons.
As for the experimental observations in Ref.~\cite{Ghigo2018}, the sudden jump in the $\lambda_L(T)$ dependence after the first dose of protons and the appearance of the long tails near $T_c$ signifying a non-mean-field behavior emphasize the complexity of the studied system.

All the mentioned issues call for the formulation of the strict signs of the $s_\pm \to s_{++}$ transition, which would allow to unambiguously assert its presence. Here we theoretically analyze the low-temperature behavior of the London penetration depth within the two-band model for the `dirty' superconductor. We derive the unmistakable signs of the transition from $s_\pm$ to $s_{++}$ state, detection of which would point towards existence of the transition.

%
%

\section{Model and approach}
Except for the extreme hole and electron dopings, the Fermi surface of most of iron-based materials consists of two or three hole sheets around the $\Gamma=(0,0)$ point and two electron sheets around the $M=(\pi,\pi)$ point of the two-iron Brillouin zone. Scattering between them with the large wave vector results in the enhanced antiferromagnetic fluctuations, which lead to the pairing with the order parameter that change sign between electron and hole pockets -- the so-called $s_\pm$ state~\cite{MazinReview,HirschfeldKorshunov2011,Korshunov2014eng}. Alternatively, the orbital fluctuations may lead to the sign-preserving $s_\pp$ state~\cite{Kontani,Bang2009,Onari,Onari2012,Yamakawa2017}. Since most experimental data including observation of a spin-resonance peak in the inelastic neutron scattering and the quasiparticle interference in tunneling experiments are in favour of the $s_\pm$ scenario~\cite{HirschfeldKorshunov2011,Hirschfeld2016}, later we consider the pairing due to spin fluctuations.

Impurity scattering in the multiband system is much more complicated than in the single band case~\cite{Golubov1997,Ummarino2007,Senga2008,Onari2009,EfremovKorshunov2011,Efremov2013,WangImp2013,KorshunovMagn2014}. One of the conclusions was that the system having the $s_\pm$ state in the clean case may preserve a finite $T_c$ in the presence of nonmagnetic disorder due to the transition to the $s_\pp$ state. It was obtained both in the strong-coupling $\Tmat$-matrix approximation~\cite{EfremovKorshunov2011} and via a numerical solution of the Bogoliubov-de Gennes equations~\cite{Yao2012,Chen2013}.

Topology of the Fermi surface in iron-based materials makes it sensible to use a two-band model as a compromise between simplicity and possibility to capture the essential physics. Previously, we have studied the $s_\pm \to s_\pp$ transition in such a model and shown that the transition can take place only in systems with the effective intraband pairing interaction present~\cite{EfremovKorshunov2011}. Physical reason for the transition is quite transparent, namely, if one of the two competing superconducting interactions leads to the state robust against impurity scattering, then although it was subdominating in the clean limit, it should become dominating while the other state is destroyed by the impurity scattering~\cite{KorshunovUFN2016}.

Here we use the two-band model~\cite{EfremovKorshunov2011,KorshunovUFN2016,ShestakovKorshunovSymmetry2018} with the following Hamiltonian,
\begin{equation}
H = \sum_{\kv, \alpha, \sigma}\xi_{\kv,\alpha}c_{\kv\alpha\sigma}^{\dagger}c_{\kv\alpha\sigma} + \sum_{\mathbf{R}_i, \sigma, \alpha, \beta}{\mathcal{U}_{\mathbf{R}_i}^{\alpha \beta}c_{\mathbf{R}_i\alpha\sigma}^{\dagger}c_{\mathbf{R}_i\beta\sigma} } + H_{SC}, \label{twoBandHimp}
\end{equation}
where the operator $c_{\kv\alpha\sigma}^{\dagger}$($c_{\kv\alpha\sigma}$) creates (annihilates) a quasiparticle with the band index $\alpha = (a,b)$, momentum $\kv$, and spin $\sigma$; $\xi_{\kv,\alpha} = {\mathbf{v}}_{F\alpha}({\kv} - {\kv}_{F\alpha})$ is the dispersion of quasiparticles linearized near the Fermi level, with ${\mathbf{v}}_{F\alpha}$ and ${\kv}_{F\alpha}$ being the Fermi velocity and Fermi momentum of the band $\alpha$, respectively. The second term in the Hamiltonian contains the impurity potential ${\mathcal{U}}_{\mathbf{R}_i}$ at a site ${\mathbf{R}_i}$, while the last term, whose exact form is not important for the current discussion, is responsible for the superconductivity. We assume that the superconducting pairing is provided by the exchange of spin fluctuations (repulsive interaction) and may include some attractive interaction (for example, electron-phonon coupling). All these interactions enter our theory through the normal and anomalous self-energy parts, which depend on the matrix of coupling constants $\hat{\Lambda}$, see Ref.~\cite{KorshunovUFN2016} for details.

The presence of nonmagnetic disorder is considered within the Eliashberg approach for multiband superconductors~\cite{allen}. To simplify the calculations, we use the quasiclassical $\xi$-integrated Green's functions,
\begin{equation}
\hat{\mathbf{g}}(\omega_n) = \left( \begin{array}{cc} \hat{\mathrm{g}}_{an} & 0 \\ 0 & \hat{\mathrm{g}}_{bn} \end{array} \right),
\label{xiIntegrated}
\end{equation}	
where $\omega_n = (2n+1)\pi T$ is the Matsubara frequency, and
\begin{equation}
\hat{\mathrm{g}}_{\alpha n} = g_{0\alpha n}\hat{\tau}_0 \otimes \hat{\sigma}_0 + g_{2\alpha n}\hat{\tau}_2 \otimes \hat{\sigma}_2, \label{gAlpha}
\end{equation}
Here, $\hat{\tau}_i$ and $\hat{\sigma}_i$ are the Pauli matrices corresponding to Nambu and spin spaces, respectively; $g_{0\alpha n}$ and $g_{2\alpha n}$ are the normal and anomalous (Gor'kov) $\xi$-integrated Green's functions in the Nambu representation,
%
\begin{equation}
g_{0 \alpha n} = -\frac{ \ii \pi N_{\alpha} \tilde{\omega}_{\alpha n} }{ \sqrt{\tilde{\omega}_{\alpha n}^2 + \tilde{\phi}_{\alpha n}^2} },
g_{2 \alpha n} = -\frac{ \pi N_{\alpha} \tilde{\phi}_{\alpha n} }{ \sqrt{\tilde{\omega}_{\alpha n}^2 + \tilde{\phi}_{\alpha n}^2} }, \label{g0g2}
\end{equation}
which depend on the density of states per spin at the Fermi level of the corresponding band ($N_{a,b}$), and on the order parameter $\tilde{\phi}_{\alpha n}$ and frequency $\tilde{\omega}_{\alpha n}$ renormalized by the self-energy.
The order parameter $\tilde{\phi}_{\alpha n}$ is connected to the gap function $\Delta_{\alpha n}$ via the renormalization factor $Z_{\alpha n} = \tilde{\omega}_{\alpha n} / \omega_{n}$, i.e.
\begin{equation}
\Delta_{\alpha n} = \tilde{\phi}_{\alpha n} / Z_{\alpha n}.
\label{eq:Delta}
\end{equation}
The impurity part of self-energy $\hat{\mathbf{\Sigma}}^\mathrm{imp}$ is calculated in the noncrossing diagrammatic approximation described by the $\mathcal{T}$-matrix approximation with the following equation,
\begin{equation}
\hat{\mathbf{\Sigma}}^\mathrm{imp}(\omega_n) = n_\mathrm{imp}\hat{\mathbf{U}} + \hat{\mathbf{U}}\hat{\mathbf{g}}(\omega_n)\hat{\mathbf{\Sigma}}^\mathrm{imp}(\omega_n), \label{Tmatrix}
\end{equation}
where $n_{\mathrm{imp}}$ is the concentration of impurities, $\hat{\mathbf{U}} = \mathbf{U} \otimes \hat{\tau}_3$, is the matrix of the impurity potential $\left( \mathbf{U} \right)_{\alpha \beta} = \mathcal{U}_{\mathbf{R}_i}^{\alpha \beta}$, consisting of intra- and interband parts $\left( \mathbf{U} \right)_{\alpha \beta} = \left( v - u \right)\delta_{\alpha \beta} + u$. The relation between the intra- and interband impurity scattering is set by a parameter $\eta = v / u$. Without loss of generality we set $\mathbf{R}_i = 0$.

It is convenient to introduce the generalized cross-section parameter
\begin{equation}
\sigma = \frac{\pi^2 N_a N_b u^2}{1 + \pi^2 N_a N_b u^2} \to
\left\{
    \begin{array}{l}
    0, \mathrm{Born~limit}, \\ 1, \mathrm{unitary~limit}
    \end{array}
  \right.
\end{equation}
and the impurity scattering rate
\begin{eqnarray}
\Gamma_{a(b)} &= 2 n_\mathrm{imp} \pi N_{b(a)} u^2 \left(1-\sigma\right) \\ \nonumber
&= \frac{2 n_\mathrm{imp} \sigma}{\pi N_{a(b)}} \to
\left\{
    \begin{array}{l}
    2n_\mathrm{imp}\pi N_{b(a)} u^2, \mathrm{Born~limit}, \\ \frac{2n_\mathrm{imp}}{\pi N_{a(b)}}, \mathrm{unitary~limit}
    \end{array}
  \right.,
\label{GammaAB}
\end{eqnarray}
For $\sigma$ and $\Gamma_{\alpha}$, there are two limiting cases: (i) Born limit corresponding to the weak impurity potential ($\pi u N_{a(b)} \ll 1$), and (ii) unitary limit corresponding to strong impurity scattering ($\pi u N_{a(b)} \gg 1$).

In the local limit, the London magnetic field penetration depth is related to the imaginary part of the optical conductivity $\sigma^{x x'}(\omega, \q = 0)$ at zero momentum $\q$ (in the local, i.e., London, limit),
\begin{equation}
 \frac{1}{\lambda_{L, x x'}^2} = \lim_{\omega \to 0} \frac{4 \pi \omega}{c^2} \mathrm{Im}~\sigma^{x x'}(\omega, \q = 0),
 \label{lambda_conductivity}
\end{equation}
where $x$ and $x'$ are axes directions of the Cartesian coordinates and $c$ is the velocity of light. If we neglect the effects of strong coupling and, in general, Fermi-liquid effects, then for the clean uniform superconductor at zero temperature we have $1 / \lambda_{L} = \sum_{\alpha}\omega_{p\alpha} / c \equiv \sum_{\alpha}\omega_{p\alpha}^{x x} / c$, where $\omega_{p\alpha}^{x x'} = \sqrt{ 8 \pi e^2 N_\alpha(0) \la v_{F\alpha}^{x} v_{F\alpha}^{x'} \ra}$ is the electron plasma frequency for the band $\alpha$, $N_\alpha(0)$ is the density of states at the Fermi level, and $v_{F\alpha}^{x}$ is the $x$-component of the Fermi velocity. For impurity scattering, vertex corrections from noncrossing diagrams vanish due to the $\q = 0$ condition. Thus, penetration depth for the multiband system can be calculated via the following expression~\cite{KorshunovUFN2016},
\begin{equation}
 \frac{1}{\lambda_{L}^{2}} = \sum_{\alpha}\left( \frac{\omega_{p\alpha}}{c} \right)^2T\sum_{n}\frac{g_{2\alpha n}^2}{\pi N_{\alpha}^2\sqrt{\tilde{\omega}_{\alpha n}^2+\tilde{\phi}_{\alpha n}^2}}. \label{lambda_squre}
\end{equation}
%
In the experiments, along with $\lambda_{L}$ the following quantities are measured: the temperature variation of the penetration depth
\begin{equation}
\Delta \lambda_{L}(T) = \lambda_{L}(T)-\lambda_{L}(0), \label{Delta_lambda}
\end{equation}
and the so-called `superfluid density'
\begin{equation}
\rho_{s}(T) = \frac{\lambda_{L}^{2}(0)}{\lambda_{L}^{2}(T)}. \label{ro_s}
\end{equation}

\section{Results}
The calculations are done in the intermediate impurity scattering limit, $\sigma = 0.5$. For simplicity, we exclude the \textit{intraband} impurity scattering assuming $\eta = 0$ and set $u=0.7$ and $v=0$. Plasma frequencies for two bands are taken from DFT (density functional theory) calculations~\cite{CharnukhaNatComm2011,YareskoPlasma2018} and are equal to $\omega_{pa}=2.34$~eV and $\omega_{pb}=1.25$~eV. The values are typical for iron pnictides, see Ref.~\cite{Charnukha2011}.

The behaviour of the $s_{\pm}$ state heavily depends on the sign of a coupling constant averaged over the bands $\langle \Lambda \rangle$. Latter is calculated using the following equation
\begin{equation}
 \langle \Lambda \rangle = \left( \Lambda_{aa} + \Lambda_{ab} \right)\frac{N_{a}}{N} + \left( \Lambda_{ba} + \Lambda_{bb} \right)\frac{N_{b}}{N},
\label{averaged_lambda}
\end{equation}
where $N = N_a+N_b$ is the total density of states. The ratio $N_a/N_b = 1/2$ and the spin fluctuation spectrum are the same as considered in the earlier studies~\cite{EfremovKorshunov2011,KorshunovMagn2014,KorshunovUFN2016}. The same is true for the matrix of the coupling constants
$\hat{\Lambda} = \left(\begin{array}{rr}
  \Lambda_{aa} & \Lambda_{ab} \\
  \Lambda_{ba} & \Lambda_{bb}
\end{array}\right)$
with the following elements:
$\hat{\Lambda} = \left(\begin{array}{rr}
    3 & -0.2 \\
    -0.1 & 0.5 \\
  \end{array}\right)$
for the $s_{\pm}$ order parameter with $\langle \Lambda \rangle > 0$ and the critical temperature $T_{c0}=41.4$~K,
$\hat{\Lambda} = \left(\begin{array}{rr}
    2 & -2 \\
    -1 & 1 \\
  \end{array}\right)$
for the $s_{\pm}$ order parameter with $\langle \Lambda \rangle < 0$ and $T_{c0}=39$~K, and
$\hat{\Lambda} = \left(\begin{array}{rr}
    3 & 0.2 \\
    0.1 & 0.5 \\
  \end{array}\right)$
for $s_{++}$ superconductor with $T_{c0}=41.4$~K.

It was shown before~\cite{ShestakovKorshunovSymmetry2018}, that the scattering rate $\Gamma_a^{crit}$ at which the transition between $s_{\pm}$ and $s_{++}$ states takes place is temperature-dependent. In the intermediate scattering limit, $\sigma=0.5$, for $T=0.01T_{c0}$ we have $\Gamma_a^{crit} = 1.15T_{c0}$.

\subsection{Penetration depth $\lambda_{L}$ and $\lambda_{L}^{-2}$ in a wide temperature range}
\begin{figure}[h]
\centering
\includegraphics[width=0.5\textwidth]{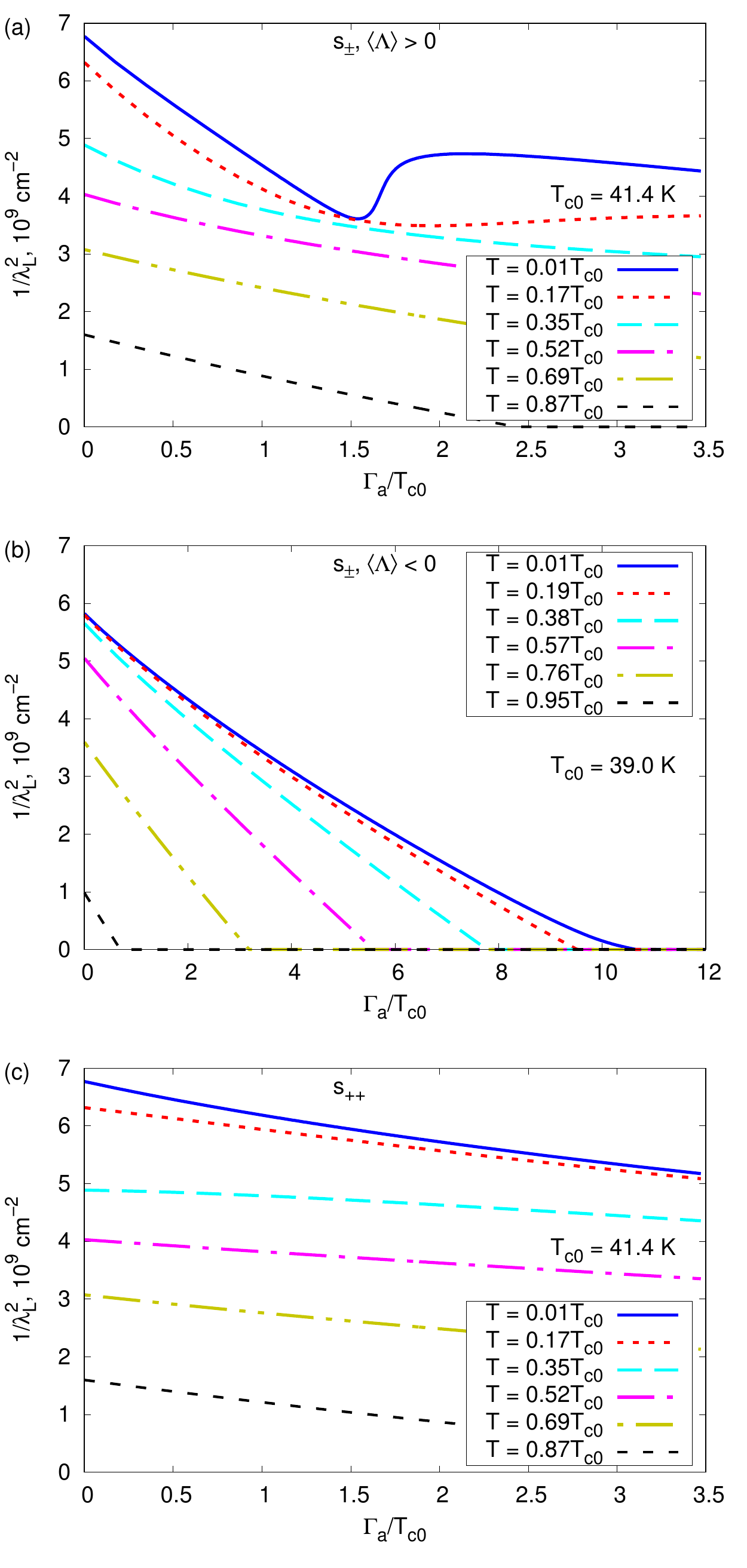}
\caption{Inverse square of the penetration depth, $\lambda_{L}^{-2}(\Gamma_a)$, at temperatures $0.01T_{c0} < T < T_{c0}$ for the $s_{\pm}$ state with $\langle \Lambda \rangle > 0$ (a) and $\langle \Lambda \rangle < 0$ (b), and for the $s_{++}$ state (c). Here, $\sigma = 0.5$, $v = 0$. $\Gamma_a$ and $T$ are normalized by $T_{c0}$. For $T=0.01T_{c0}$ and $T=0.02T_{c0}$, the graphs almost overlap.}
\label{fig_lambda_sqr_T}
\end{figure}

%
%
Since the optical conductivity, and thus the inverse square of the penetration depth is the response function, see Eq.~(\ref{lambda_conductivity}), it is $\lambda_{L}^{-2}$ that can be directly observed and measured, not $\lambda_{L}$ itself. Therefore, from the point of view of experimental detection of the transition, it turns out to be more convenient to consider $\lambda_{L}^{-2}$ instead of $\lambda_{L}$. In Fig.~\ref{fig_lambda_sqr_T}(a), the $s_{\pm} \to s_{++}$ transition at $T=0.01-0.02T_{c0}$ manifests itself as a jump in $\lambda_{L}^{-2}(\Gamma_a)$. There is a well pronounced minimum at the transition point. Note that the minimum can not be eliminated by changing the scale of the graph. For the $s_{++}$ gap, Fig.~\ref{fig_lambda_sqr_T}(c), $\lambda_{L}^{-2}$ slightly decreases as a function of $\Gamma_a$ retaining a finite value below $T_{c}$, while for the $s_{\pm}$ gap with $\langle \Lambda \rangle < 0$, Fig.~\ref{fig_lambda_sqr_T}(b), inverse square of the penetration depth vanishes rapidly at any temperature.

\begin{figure}[h]
\centering
\includegraphics[width=0.5\textwidth]{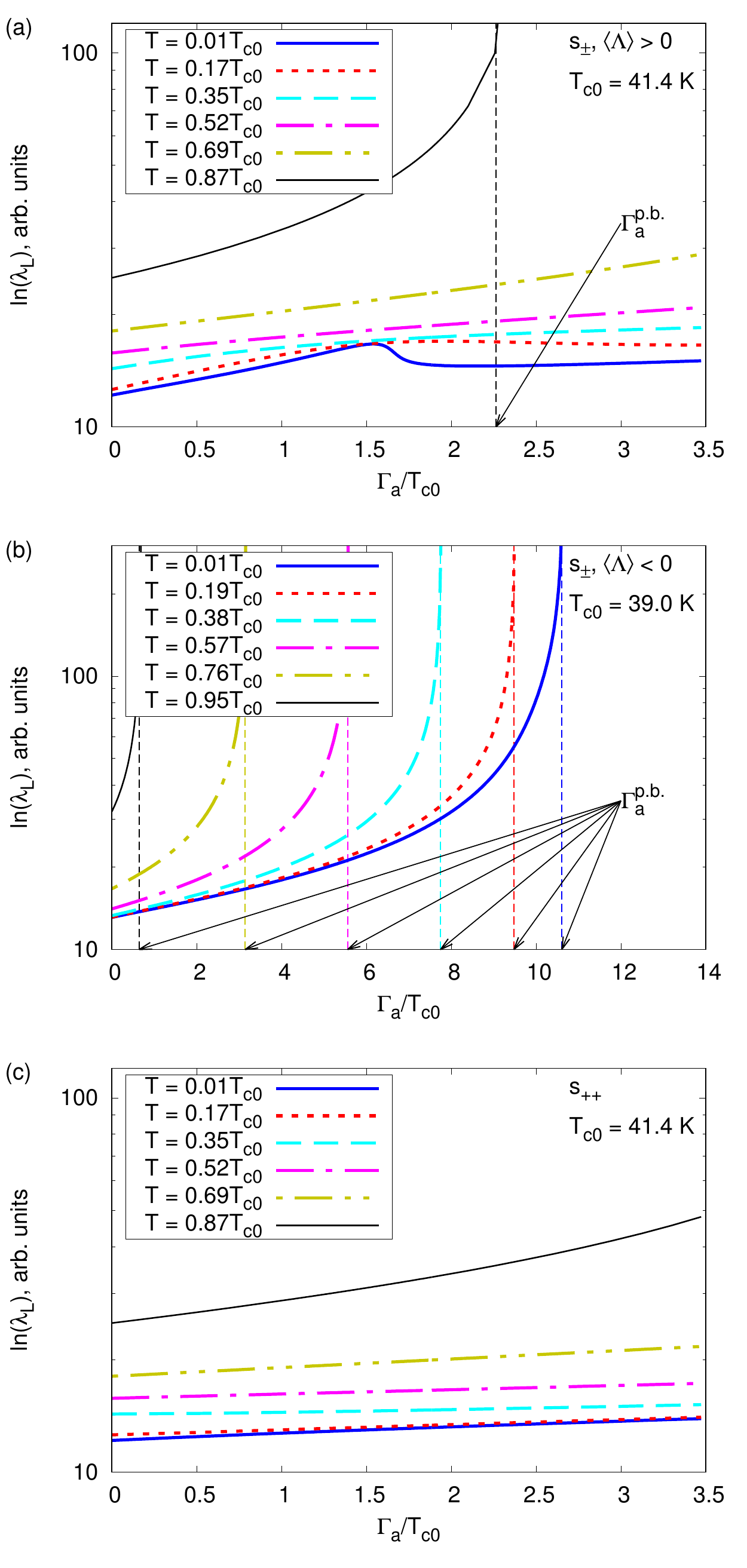}
\caption{Log plot of the penetration depth $\lambda_{L}(\Gamma_a)$ at various temperatures $0.01T_{c0} < T < T_{c0}$ for the $s_{\pm}$ state with $\langle \Lambda \rangle > 0$ (a) and with $\langle \Lambda \rangle < 0$ (b), and for the $s_{++}$ state (c). Here, $\sigma = 0.5$, $v = 0$. $\Gamma_a$ and $T$ are normalized by $T_{c0}$. For $T=0.01T_{c0}$ and $T=0.02T_{c0}$ the graphs almost overlap.}
\label{fig_lambda_T}
\end{figure}

At any temperature above approximately $0.02T_{c0}$, the $s_{\pm} \to s_{++}$ transition is not very pronounced, see Fig.~\ref{fig_lambda_T}(a). Moreover, the behaviour of $\lambda_{L}$ for the $s_{\pm}$ state with $\langle \Lambda \rangle > 0$ and the $s_{++}$ state are similar 
up to $T \approx 0.8T_{c0}$, compare Figs.~\ref{fig_lambda_T}(a) and (c). At higher temperatures, however, $\lambda_{L}$ increases faster as a function of $\Gamma_a$ in the former case. It happens because $T_{c}$ in the $s_{++}$ state emerging after the $s_{\pm} \to s_{++}$ transition appears to be suppressed more intensively than the `genuine' $s_{++}$ state shown in Fig.~\ref{fig_lambda_T}(c). And at $T > 0.8T_{c0}$, there is a range of impurity scattering rates starting from $\Gamma_a^{\mathrm{p.b.}}$ (`pair breaking') for which the superconductivity is fully suppressed. In Fig.~\ref{fig_lambda_T}(a), it is shown for $T=0.87T_{c0}$. In the case of the $s_{\pm}$ state with $\langle \Lambda \rangle < 0$, nonmagnetic disorder destroys the superconducting state at any temperature within the range of $0.01T_{c0} < T < T_{c0}$, which is indicated by the presence of $\Gamma_a^{\mathrm{p.b.}}$ at each temperature, see Fig.~\ref{fig_lambda_T}(b).

\subsection{Penetration depth at a minimal temperature $T \to 0$}
Because we use Matsubara technique in our calculations, we can not consider exact zero temperature case. Apparently, this is in line with the experimental situation where one can not make measurements at $T=0$. Therefore, we set $\lambda_{L}(0) \approx \lambda_{L}(T=T_{min})$ with $T_{min} = 0.01T_{c0}$.

\begin{figure}[t]
\centering
\includegraphics[width=0.5\textwidth]{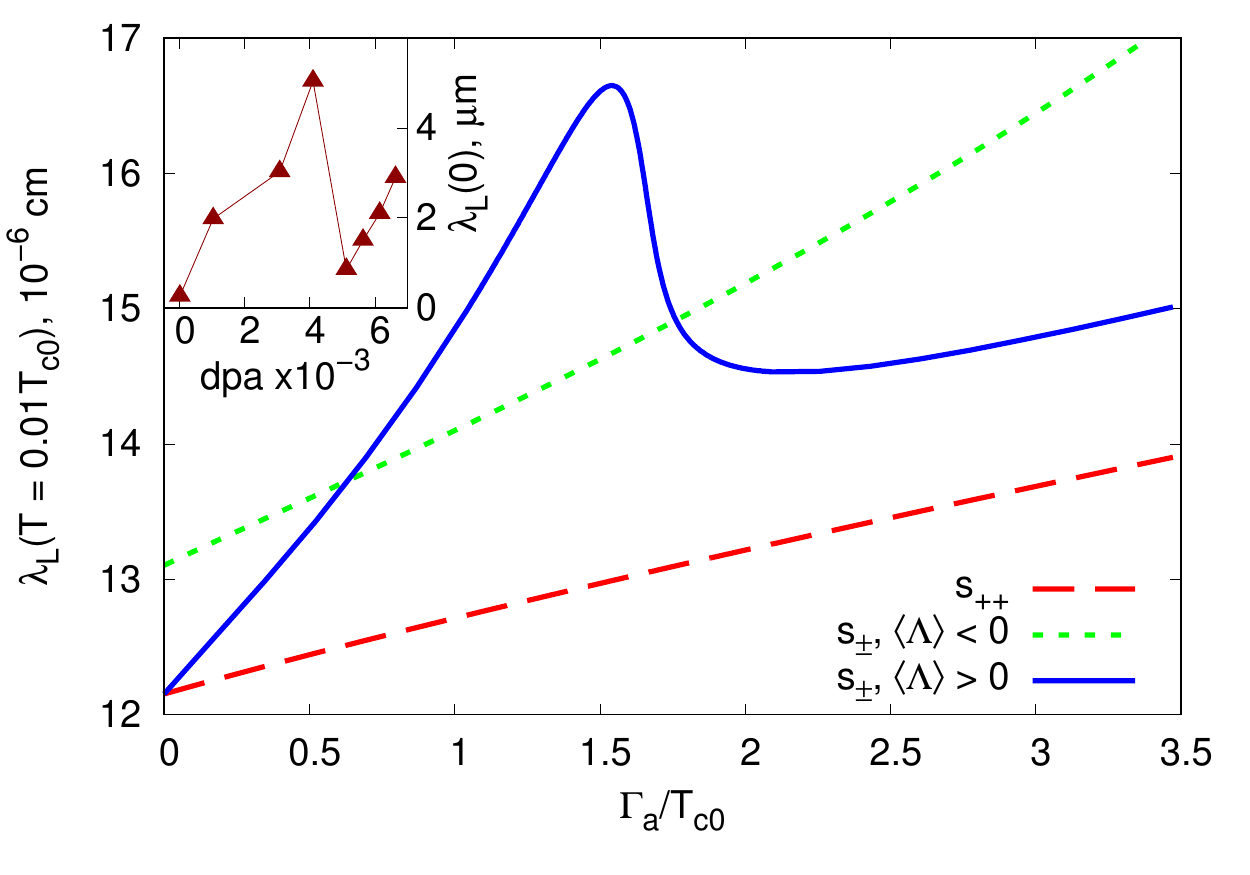}
\caption{Penetration depth $\lambda_{L}(\Gamma_a)$ at temperature $T = 0.01T_{c0}$ for the $s_{\pm}$ state with $\langle \Lambda \rangle > 0$ and with $\langle \Lambda \rangle < 0$, and for the $s_{++}$ superconductor. Here, $\sigma = 0.5$, $v = 0$. $\Gamma_a$ and $T$ are normalized by $T_{c0}$. Data points from Ref.~\cite{Ghigo2018} are shown in the inset to demonstrate the nonmonotonic experimental dependence of $\lambda_{L}(T \to 0)$ and qualitative similarity with our result for the $s_{\pm}$ state with $\langle \Lambda \rangle > 0$.}
\label{fig_lambda_minT}
\end{figure}

As is seen in Fig.~\ref{fig_lambda_minT} (blue curve), the transition between the $s_{\pm}$ and $s_{++}$ superconducting states is manifested in the penetration depth at $T \to 0$ as a sharp change in the dependence of $\lambda_{L}$ on the impurity scattering rate. Before the transition, $\lambda_{L}(\Gamma_a)$ rapidly increases indicating the suppression of the $s_{\pm}$ gap. The similar behavior is obtained for the $s_{\pm}$ state with $\langle \Lambda \rangle < 0$, see the green curve in Fig.~\ref{fig_lambda_minT}. After the transition, which can be seen as an abrupt jump in $\lambda_{L}(\Gamma_a)$, the penetration depth increases less intensively that corresponds to the $s_{++}$ state shown by the red curve in Fig.~\ref{fig_lambda_minT}. Note that the slope of the curve before the transition is different from that after the transition, which is an additional indication of the transfer from $s_{\pm}$ to $s_{++}$ state. By and large, the blue curve in Fig.~\ref{fig_lambda_minT} reproduces qualitatively the experimental results obtained by Ghigo \textit{et al.}~\cite{Ghigo2018}, however, there are some quantitative difference.

\subsection{Temperature variation of the penetration depth $\Delta \lambda_{L}(T)$}
\begin{figure}[h]
\centering
\includegraphics[width=0.5\textwidth]{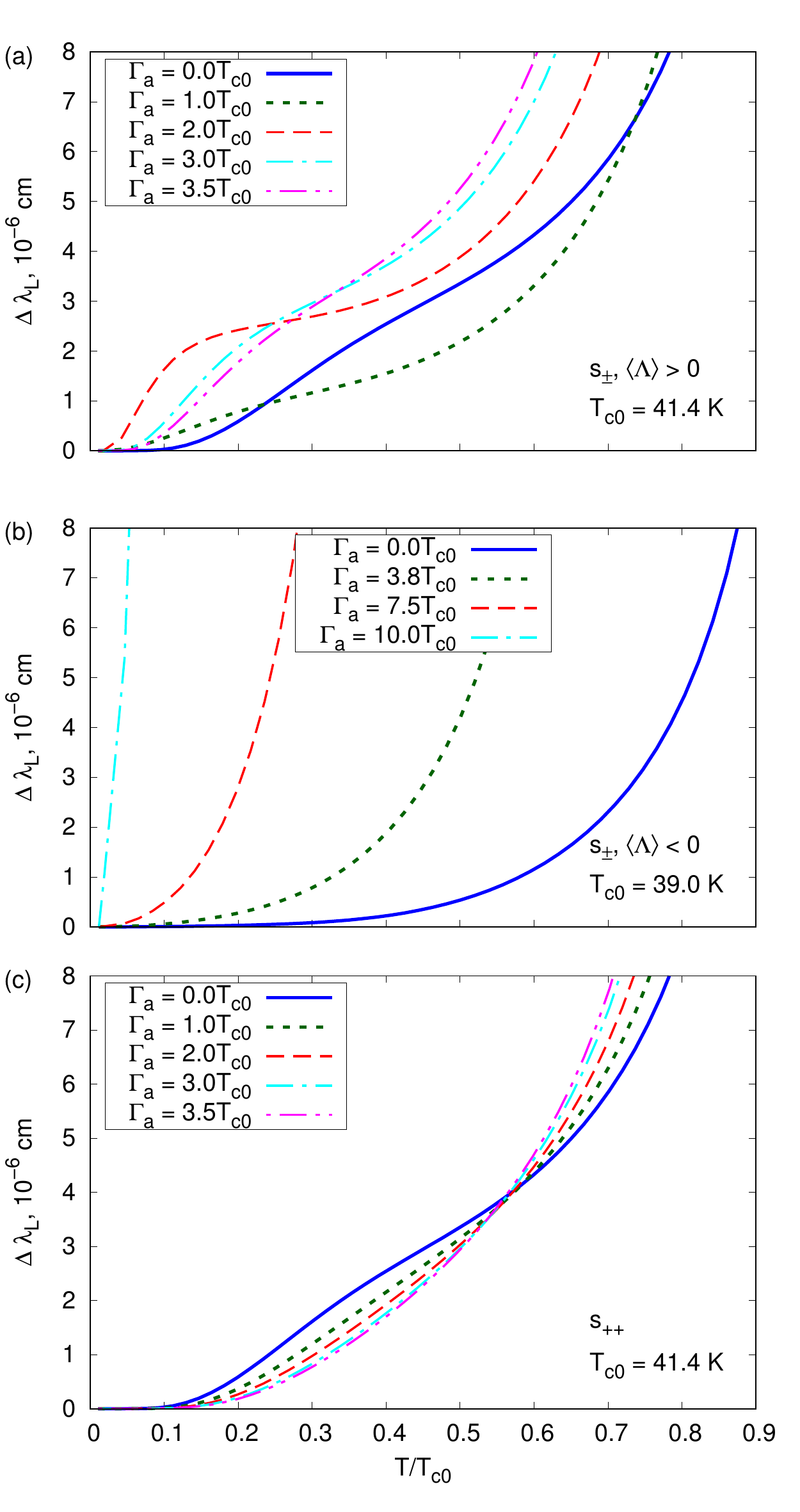}
\caption{Temperature variation of the penetration depth $\Delta \lambda_{L}(T) = \lambda_{L}(T)-\lambda_{L}(T_{min})$ for the $s_{\pm}$ state with $\langle \Lambda \rangle > 0$ (a) and with $\langle \Lambda \rangle < 0$ (b), and for the $s_{++}$ state (c). Calculations were done in the intermediate scattering limit, $\sigma=0.5$, with the interband-only  impurity potential, $v=0.0u$. $\Gamma_a$ and $T$ are normalized by $T_{c0}$.}
\label{fig_Delta_lambda}
\end{figure}

To calculate the variation of the penetration depth $\Delta \lambda_{L}(T)$ using Eq.~(\ref{Delta_lambda}) we again set $\lambda_{L}(0) \approx \lambda_{L}(T_{min})$. Dependence of $\Delta \lambda_{L}(T)$ on the impurity scattering rate $\Gamma_a$ in the $s_{++}$ and $s_{\pm}$ states are shown in Fig.~\ref{fig_Delta_lambda}. For the $s_{++}$ gap, Fig.~\ref{fig_Delta_lambda}(c), the slope of $\Delta \lambda_{L}(T)$ decreases with the increasing $\Gamma_a$, while the slope for the $s_{\pm}$ gap with $\langle \Lambda \rangle < 0$, Fig.~\ref{fig_Delta_lambda}(b), becomes steeper. In the case of the $s_{\pm}$ gap with $\langle \Lambda \rangle > 0$, Fig.~\ref{fig_Delta_lambda}(a), both of these features are present:
before the $s_{\pm} \rightarrow s_{++}$ transition the slope increases, and after the transition it decreases.

Two gaps for the $s_{++}$ state, initially (at $\Gamma_a=0$) having different values, start to change with disorder so their magnitudes become closer. Thus the initial two-gap behaviour switches to the single-gap-like dependence of $\Delta \lambda_{L}(T)$ at higher $\Gamma_a$. Such a change is the reason for the crossing of different curves around $0.6T_{c0}$ in Fig.~\ref{fig_Delta_lambda}(c).

\subsection{Temperature dependence of the superfluid density $\rho_{s}(T)$}
\begin{figure}[h]
\centering
\includegraphics[width=0.5\textwidth]{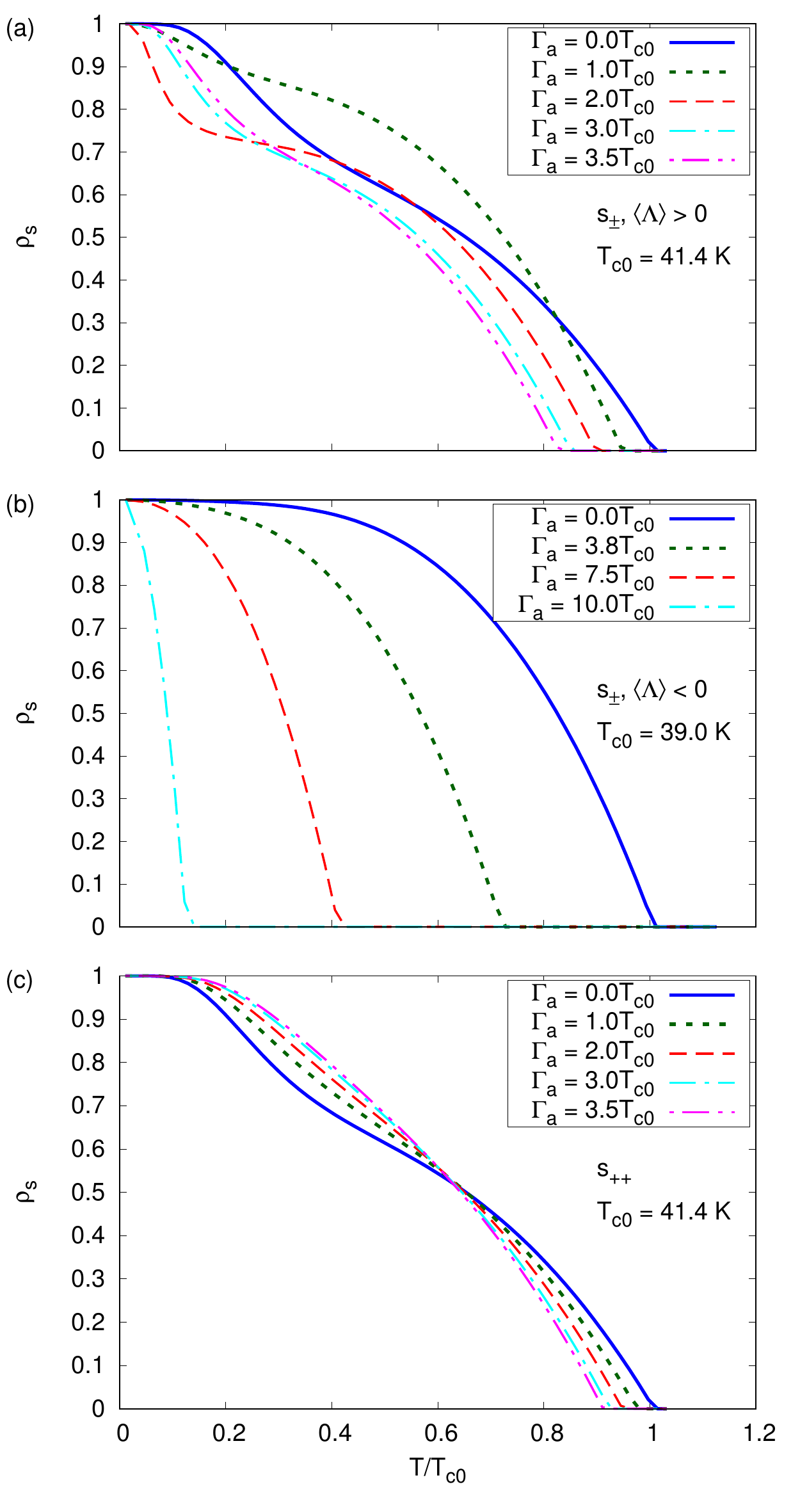}
\caption{Superfluid density $\rho_{s}$ for the $s_{\pm}$ state with $\langle \Lambda \rangle > 0$ (a) and $\langle \Lambda \rangle < 0$ (b), and for the $s_{++}$ superconductor (c). Here, $\sigma = 0.5$, $v = 0$. $\Gamma_a$ and $T$ are normalized by $T_{c0}$.}
\label{fig_ro_s}
\end{figure}

\begin{figure}[h]
\centering
\includegraphics[width=0.5\textwidth]{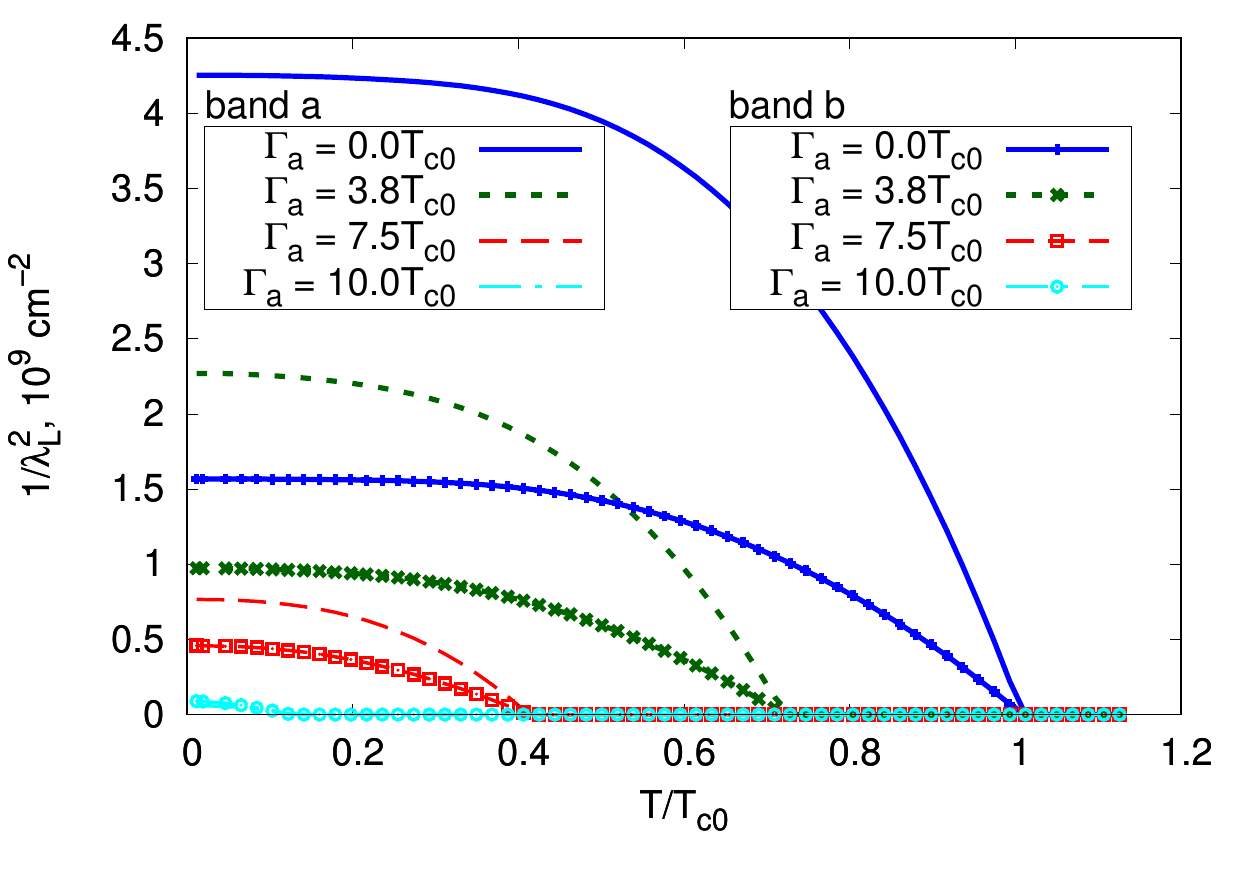}
\caption{Temperature dependence of $\lambda_{L}^{-2}$ for the $s_{\pm}$ state with $\langle \Lambda \rangle < 0$. $\Gamma_a$ and $T$ are normalized by $T_{c0}$.}
\label{fig_lambda_sqr_partial}
\end{figure}

For the pure $s_{\pm}$ superconductor with $\langle \Lambda \rangle > 0$, the temperature dependence of the superfluid density $\rho_{s}(T)$ defined by Eq.~(\ref{ro_s}) demonstrates an exemplary behaviour for a two-band superconductor having two unequal gaps, see Fig.~\ref{fig_ro_s}(a), $\Gamma_a=0$ case. Increasing the impurity scattering rate changes the two-gap behavior of $\rho_{s}$ in the vicinity of the $s_{\pm} \to s_{++}$ transition into the one that is specific for a single-gap $s$-wave superconductor. This happens because the smaller gap goes through zero while changing its sign. Further increase of the concentration of impurities restores the two-gap behavior of $\rho_{s}(T)$ that tends to a single-gap one at the highest values of $\Gamma_a$ in the $s_{++}$ state. Such a situation is quite different from the case of the $s_{\pm}$ state with $\langle \Lambda \rangle < 0$, which exhibits in our calculations the single-gap behavior of $\rho_{s}(T)$ despite the presence of two unequal gaps, see Fig.~\ref{fig_ro_s}(b). The reason for this is the qualitatively similar temperature dependence of $\lambda_{L}^{-2}$ for both gaps, see partial contributions from both bands into $\lambda_{L}^{-2}(T)$ in Fig.~\ref{fig_lambda_sqr_partial}. The behavior shown in Fig.~\ref{fig_ro_s}(a) also differs from that for the $s_{++}$ superconductor, Fig.~\ref{fig_ro_s}(c), which changes directly from the typical two-gap dependence for low scattering rates to the single-gap $\rho_{s}(T)$ for high values of $\Gamma_a$.

Based on Figs.~\ref{fig_Delta_lambda} and~\ref{fig_ro_s}, we admit that in the clean limit there is no difference between $s_{++}$ state and $s_{\pm}$ state with $\langle \Lambda \rangle > 0$ in such quantities as $\Delta \lambda_{L}$ and $\rho_{s}$.

\section{Conclusions}

Here we derived specific features of the $s_\pm \to s_{++}$ transition in the temperature and impurity scattering rate dependencies of the penetration depth. Observation of those features can serve as criteria for detection of the transition. First one is the sharp change in the dependence of the penetration depth $\lambda_{L}$ on the impurity scattering rate at $T \to 0$. Moreover, the slope of the curve before the transition is different from that after the transition, which is an additional indication of the transfer between the $s_{\pm}$ and $s_{++}$ states.

Second feature is connected to the slope of the relative change in the penetration depth, $\Delta \lambda_{L}(T) = \lambda_{L}(T)-\lambda_{L}(0)$, as a function of temperature -- before the $s_{\pm} \to s_{++}$ transition the slope increases, and after the transition it decreases.

Third feature is the sharp jump in the inverse square of the penetration depth as a function of the impurity scattering rate, $\lambda_{L}^{-2}(\Gamma_a)$, at the $s_{\pm} \to s_{++}$ transition.

And the last one is the temperature dependence of the superfluid density $\rho_{s}(T)$ that exhibits almost the single-gap behavior in the vicinity of the $s_{\pm} \to s_{++}$ transition and upon increase of the impurity scattering rate restores the two-gap behavior.

Results here are obtained in the intermediate scattering limit, $\sigma=0.5$. Changing $\sigma$ would change the exact position of the transition in the $T$-$\Gamma_a$ phase diagram, see Ref.~\cite{ShestakovKorshunovSymmetry2018}. The transition itself, however, remains in a wide range of $\sigma$'s except for the unitary limit ($\sigma=1$) with nonuniform impurity potential $\eta \neq 1$. Therefore, the discussed specific features of the penetration depth can be observed for a system exhibiting the $s_\pm \to s_{++}$ transition in a wide range of parameters.

\ack

We are grateful to D.V. Efremov, A.S. Fedorov, S.G. Ovchinnikov, E.I. Shneyder, D. Torsello, and A.N. Yaresko for useful discussions.
This work was supported in part by the Russian Foundation for Basic Research (RFBR) grant 19-32-90109 and by RFBR and Government of Krasnoyarsk Territory and Krasnoyarsk Regional Fund of Science to the Research Projects ``Electronic correlation effects and multiorbital physics in iron-based materials and cuprates'' grant 19-42-240007.

\section*{References}

\bibliographystyle{iopart-num}
\bibliography{mmkbibl8}

\end{document}